\documentclass[pra,
groupedaddress,
amsmath,
amssymb,
twocolumn,
]{revtex4-1}

\usepackage[hmarginratio=1:1,top=3cm,height=24.7cm,width=18cm,a4paper]{geometry}
\usepackage{graphicx}
\usepackage{dcolumn}
\usepackage{dsfont}
\usepackage{bm}

\setcitestyle{super}
\newcommand{\bra}[1]{\langle #1|}
\newcommand{\ket}[1]{|#1\rangle}
\newcommand{\tr}{\rm tr}
\newcommand{\ii}{\mathrm{i}}

\begin{document}

\title{Experimental Realization of Non-Abelian Geometric Gates}

\author{A.~A.~Abdumalikov,~Jr.}
\email{abdumalikov@phys.ethz.ch}
\affiliation{Department of Physics, ETH Z\"urich, CH-8093, Z\"urich, Switzerland}
\author{J.~M.~Fink}
\altaffiliation[Now at ]{Institute for Quantum Information and Matter, California Institute of Technology, Pasadena, CA 91125, USA}
\affiliation{Department of Physics, ETH Z\"urich, CH-8093, Z\"urich, Switzerland}
\author{K.~Juliusson}
\affiliation{Department of Physics, ETH Z\"urich, CH-8093, Z\"urich, Switzerland}
\author{M.~Pechal}
\affiliation{Department of Physics, ETH Z\"urich, CH-8093, Z\"urich, Switzerland}
\author{S.~Berger}
\affiliation{Department of Physics, ETH Z\"urich, CH-8093, Z\"urich, Switzerland}
\author{A.~Wallraff}
\affiliation{Department of Physics, ETH Z\"urich, CH-8093, Z\"urich, Switzerland}
\author{S.~Filipp}
\affiliation{Department of Physics, ETH Z\"urich, CH-8093, Z\"urich, Switzerland}

\date{\today}

\maketitle

\textbf{The geometric aspects of quantum mechanics are underlined most prominently by the concept of geometric phases, which are acquired whenever a quantum system evolves along a closed path in Hilbert space. The geometric phase is determined only by the shape of this path \cite{Pancharatnam1956,Mead1979a,Berry1984,Aharonov1987} and is -- in its simplest form -- a real number.
However, if the system contains degenerate energy levels, matrix-valued geometric phases, termed non-abelian holonomies, can emerge \cite{Wilczek1984}. They play an important role for the creation of synthetic gauge fields in cold atomic gases \cite{Dalibard2011} and the description of non-abelian anyon statistics \cite{Read2009}. Moreover, it has been proposed to exploit non-abelian holonomic gates for robust quantum computation \cite{Zanardi1999, Duan2001, Pachos2012}. In contrast to abelian geometric phases \cite{Berry2010}, non-abelian ones have been observed only in nuclear quadrupole  resonance experiments with a large number of spins and without fully characterizing the geometric process and its non-commutative nature \cite{Tycko1987,Zwanziger1990}. Here, we realize non-abelian holonomic quantum operations \cite{Anandan1988,Sjoqvist2012} on a single superconducting artificial three-level atom \cite{Koch2007} by applying a well controlled two-tone microwave drive. Using quantum process tomography, we determine fidelities of the resulting non-commuting gates exceeding $95\%$. We show that a sequence of two paths in Hilbert space traversed in different order yields inequivalent transformations, which is an evidence for the non-abelian character of the implemented holonomic quantum gates. In combination with two-qubit operations, they form a universal set of gates for holonomic quantum computation.}

A cyclic evolution of a non-degenerate quantum system is in general accompanied by a phase change of its wave function. The acquired abelian phase can be divided into two parts: The dynamical phase which is proportional to the evolution time and the energy of the system, and the geometric phase which depends only on the path of the system in Hilbert space. This characteristic feature leads to a resilience of the geometric phase to certain fluctuations during the evolution \cite{Leek2007,Filipp2009a,Wu2012b}, a property which has attracted particular attention in the field of quantum information processing \cite{Sjoqvist2008}. However, universal quantum computation cannot be based on simple phase gates, which modify only the relative phase of a superposition state, unless they act on specific basis states \cite{Zhu2003}. Furthermore, geometric operations acting on degenerate subspaces have been proposed for holonomic quantum computation fully based on geometric concepts \cite{Zanardi1999}. In this scheme, quantum bits are encoded in a doubly degenerate eigenspace of the system hamiltonian $h(\vec{\lambda})$. The parameters $\vec{\lambda}$ are varied to induce a cyclic evolution of the system. When the system returns back to its initial state, it can acquire not only a simple geometric phase factor, but also undergoes a path-dependent unitary transformation, a non-abelian holonomy, which causes a transition between the eigenstates in the degenerate subspace. Because these transformations depend only on the geometric properties of the path, they share the noise resilience of the geometric phase.

In the original proposal \cite{Zanardi1999}, the parameters $\vec{\lambda}$ are changed adiabatically in time to guarantee the persistence of the degeneracy.  Adiabatic holonomic gates have been proposed for trapped ions \cite{Duan2001}, superconducting qubits \cite{Faoro2003, Kamleitner2011} and semiconductor quantum dots \cite{Solinas2003}. However, they are difficult to realize in experiment because of the long evolution time needed to fulfill the adiabatic condition. Instead, Sj\"oqvist et al.~\cite{Sjoqvist2012} have proposed a scheme based on non-adiabatic non-abelian holonomies \cite{Anandan1988} which combines universality and speed and can thus be implemented in experiments.

\begin{figure}
    \includegraphics{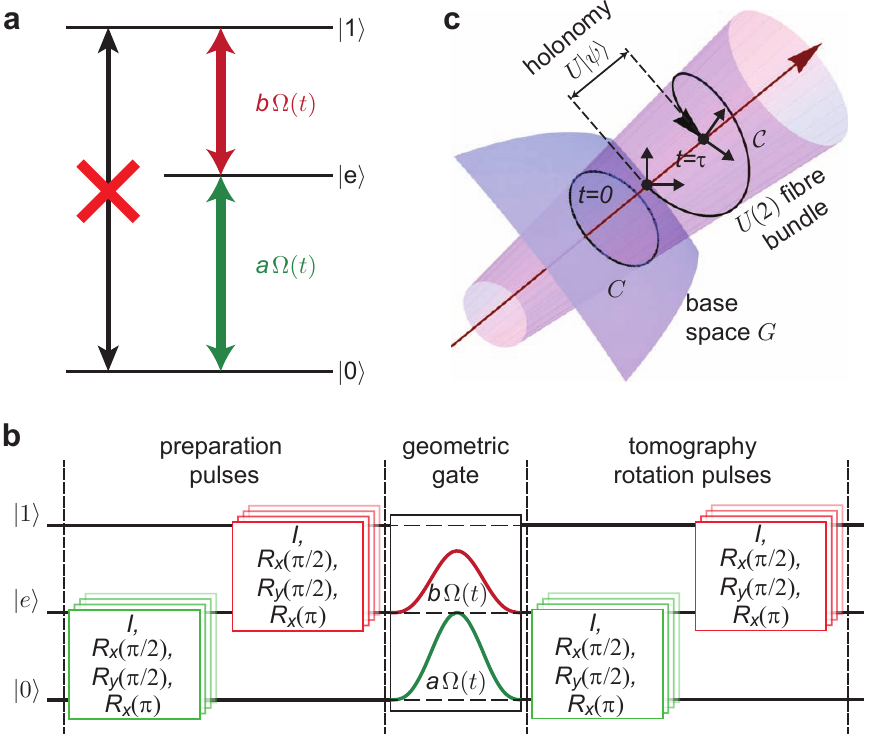}
    \caption{\textbf{Geometric gate operation on a three-level system.} \textbf{a,} Two drive tones with amplitudes $a$ and $b$ and pulse envelope $\Omega(t)$ couple the $|0\rangle$ and $|1\rangle$ to the $|e\rangle$ state. To lowest order, the direct transition between $|0\rangle$ and $|1\rangle$ states is forbidden. \textbf{b,} Pulse sequence for process tomography: The input state is prepared by sequentially applying pulses on the $|0\rangle\leftrightarrow|e\rangle$ and $|e\rangle\leftrightarrow|1\rangle$ transitions (see Methods). The holonomic gate is formed by the simultaneous application of two pulses with envelope $\Omega(t)$ and amplitudes $a$ and $b$. A set of pulses on the $|0\rangle\leftrightarrow|e\rangle$ and $|e\rangle\leftrightarrow|1\rangle$ transitions followed by the measurement completes the process tomography. \textbf{c,} Holonomic gate represented on a fibre bundle. A path in the fiber bundle connects initial and final states. The projection of the path in the fiber bundle onto the base manifold spanned by the states $\{\ket{\psi_i}\bra{\psi_i}\}$, forms a closed loop $\mathcal{C}$ along which the system is parallel transported. The difference between the initial and final points lying on a single fiber corresponds to the matrix valued holonomy $U(\mathcal{C})$, an element of the unitary group $U(N)$.}\label{fig1}
\end{figure}

The main idea is to generate a non-adiabatic and cyclic state evolution in a three-level system which results in a purely geometric operation on the degenerate subspace spanned by the computational basis states $\ket{0}$ and $\ket{1}$. The third state $\ket{e}$ acts as an auxiliary state and remains unpopulated after the gate operation. This is achieved by driving the system with two resonant microwave pulses (Fig.~\ref{fig1}a)
with identical time-dependent envelope $\Omega(t)$, but different amplitudes $a$ and $b$ satisfying $|a|^2 + |b|^2 =1$ (see Fig.~\ref{fig1}c). The hamiltonian of the system in the interaction picture is
\begin{equation}
h(t)=\frac{\hbar\Omega(t)}{2}(a\ket{e}\bra{0}+b\ket{e}\bra{1}+h.c.),
\end{equation}
and it causes the initial basis vectors to evolve to the states $\ket{\psi_i(t)} = T{\exp}\left(-\frac{\ii}{\hbar}\int\limits_0^th(t')dt'\right)\ket{i}$ $(i,j = 0,1)$, where $T$ denotes time ordering. In contrast to adiabatic schemes, the $\ket{\psi_i(t)}$ are not instantaneous eigenstates of $h(t)$. 
By keeping the complex amplitude ratio $a/b$ of the pulses constant, no transitions between states are induced and the evolution satisfies the parallel transport condition, $\bra{\psi_i(t)}h(t)\ket{\psi_j(t)}=0$. As a consequence, the evolution is purely geometric with no dynamic contributions.  If the pulse length $\tau$ is chosen such that $\int\limits_0^\tau\Omega(t)dt=2\pi$, the degenerate subspace undergoes a cyclic evolution, and the final operator which acts on the $\ket{0}$ and $\ket{1}$ basis states is
\begin{equation}\label{idealproc}
U(\mathcal{C})=\left(
\begin{array}{cc}
\cos\theta & e^{\ii\phi}\sin\theta\\
e^{-\ii\phi}\sin\theta & -\cos\theta\\
\end{array}
\right),
\end{equation}
where we have parameterized the drive amplitudes by the relation $e^{\ii\phi}\tan\theta/2=a/b$.
In a geometric picture, the dynamics of the system can be visualized on a fibre bundle in which the closed loop $\mathcal{C}$ in the base manifold determines the holonomic operation $U(\mathcal{C})$ (Fig.~\ref{fig1}c).
Different values of $\theta$ and $\phi$ correspond to different paths $\mathcal{C}$.

\begin{figure}[t]
    \includegraphics{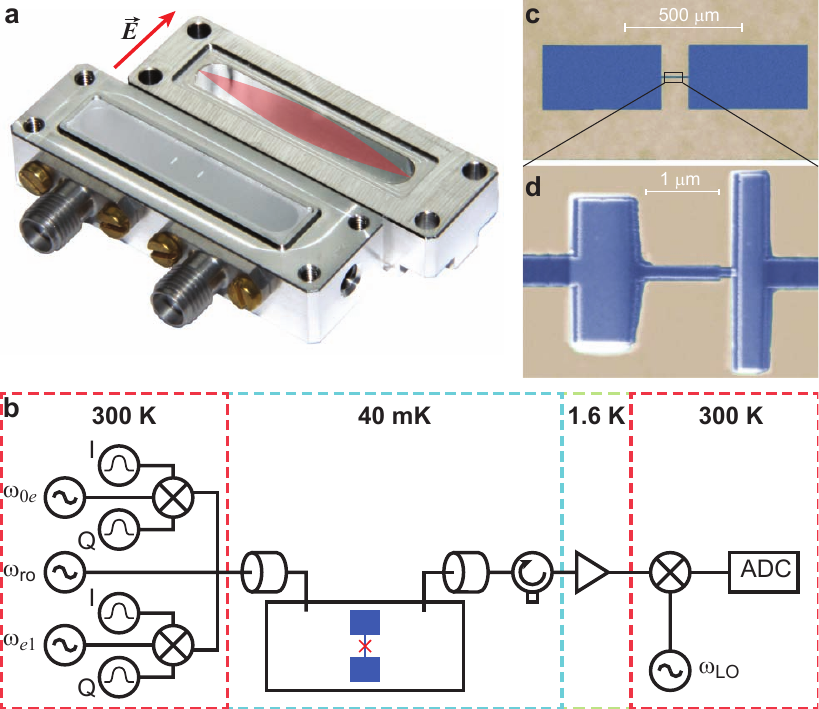}
    \caption{\textbf{Transmon qubit in a cavity resonator.} \textbf{a,} Aluminium cavity with an embedded sapphire chip containing two transmon qubits. The left qubit is not used in the experiment. The electric field profile $\vec{E}$ of the fundamental mode is sketched in the upper part of the cavity. \textbf{b}, Microwave control pulses at the transition frequencies $\omega_{0e}$ and $\omega_{e1}$ are created by modulating two continuous  signals using in-phase/quadrature $(I/Q)$ mixers. Modulation pulses in the $I$ and $Q$ channels are generated using arbitrary waveform generators. The control pulses and the readout pulse at frequency $\omega_{\rm ro}$ are combined and fed into the cavity. The transmitted signal is amplified and detected in a heterodyne measurement at room temperature and analyzed on a computer using a digitizer (ADC).  \textbf{c,} Optical and \textbf{d} scanning electron micrograph of the employed transmon device.}\label{fig2}
\end{figure}

In our experiments, we realize the holonomic gates using a three-level superconducting artificial atom of the transmon type embedded in a 3D cavity \cite{Paik2011} (Fig.\,\ref{fig2}a).
The cavity is made of aluminium with inner dimensions $32\times15.5\times5$\,mm$^3$. The frequency of the fundamental mode is $\omega_{\rm ro}/2\pi\simeq8.999$\,GHz as measured by transmission spectroscopy using the circuit shown in Fig.~\ref{fig2}b. The transmon is made of two $500\times250\,\mu$m$^2$ aluminium electrodes separated by $130\,\mu$m and connected via a Josephson junction (Fig.\,\ref{fig2}c,d). It is oriented parallel to the electric field of the fundamental mode pointing along the smallest dimension of the cavity (Fig.\,\ref{fig2}a). The measured coupling strength between the transmon and the cavity field is $g/2\pi\approx110$\,MHz. To read out the state of the transmon we measure the state-dependent transmission through the cavity \cite{Bianchetti2009}. The transition frequencies measured by Ramsey spectroscopy are $\omega_{0e}/2\pi\approx8.086$\,GHz and $\omega_{e1}/2\pi\simeq7.776$\,GHz. The decay times of both excited states are $T_{1}=7\pm0.1\,\mu$s, and the dephasing times are $T_2^{0e}=8.0\pm0.1\,\mu s$ and  $T_2^{e1}=3.9\pm0.1\,\mu s$.

\begin{figure}[t]
    \includegraphics{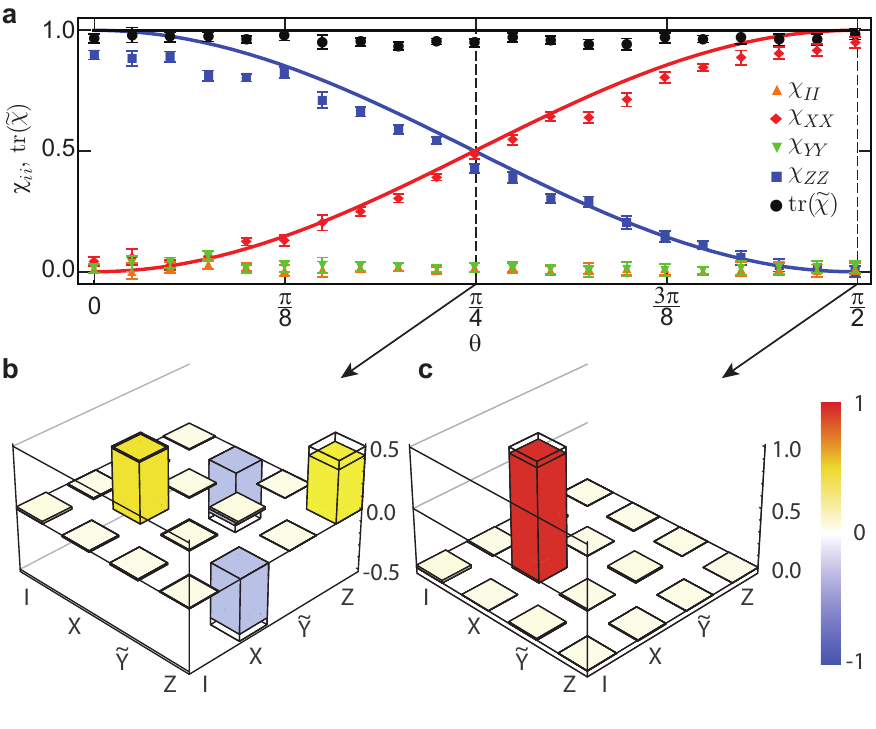}
    \caption{\textbf{Process tomography of holonomic gates.} \textbf{a,} Diagonal elements $\chi_{ii}$ of the     process matrices for ideal (lines) and experimental (dots) geometric gates as a function of $\theta$. Black circles correspond to the trace of the reduced process matrix $\tilde\chi$. (\textbf{b}) Bar chart of the real part of the reduced measured process matrix $\tilde\chi_{exp}$ of the geometric Hadamard gate $H$  and (\textbf{c}) the $NOT$ gate . The wire frames show the theoretically expected values. The small (0.3\%) imaginary parts of $\tilde\chi$ are not shown.}\label{fig3}
\end{figure}

Different holonomic gates are realized by adjusting the ratio of the two drives $a/b=e^{\ii\phi}\tan\theta/2$ to values between $0$ and $1/\sqrt{2}$ with $\phi=0$. In geometric terms, this corresponds to a gradual deformation of the loop $\mathcal{C}$ in the projective Hilbert space resulting in a continuous change from the phase-flip gate $\sigma_z$ to the $NOT$ gate $\sigma_x$. The envelopes $\Omega(t)$ are truncated Gaussian pulses \cite{Motzoi2009} with a width of $\sigma=10$\,ns and a total pulse length of 40\,ns. With pulses of this length, off-resonant driving of neighboring transitions is negligible. The performance of the holonomic gates is characterized by process tomography performed on all three levels  (see Methods). The diagonal elements of the reduced process matrix $\tilde\chi$ are shown in Fig.\,\ref{fig3}a as a function of $\theta$ with $\phi=\pi$.
The experimentally obtained results are in good agreement with theory. For $\theta=0$ a single drive on the $\ket{e}\leftrightarrow\ket{1}$ transition causes a phase shift corresponding to $\chi_{ZZ}=1$ and the operation $U(\theta=0)=\sigma_z$. In the case $\theta=\pi/4$, the Hadamard transformation $H=(\sigma_z-\sigma_x)/\sqrt{2}$ is generated (Fig.\,\ref{fig3}b). $\theta=\pi/2$ corresponds to pulses with equal amplitude and generates a $NOT$ gate $\sigma_x$ (Fig.\,\ref{fig3}c). Because of dephasing and relaxation of both excited states as well as finite fidelities of the microwave pulses, the $\ket{e}$ state is slightly populated after the gate operation. This leakage is quantified by computing the trace $\mathrm{tr}\left(\tilde\chi\right)\approx0.96$ of the reduced process matrix $\tilde\chi$ (black dots).

The fidelity of the Hadamard transformation is ${\mathcal F}_{H}\approx95.4\%$ and the fidelity of the $NOT$ operation is ${\cal F}_{NOT}\approx97.5\%$. The numerical solution of a master equation including dissipative processes results in fidelities of ${\mathcal F}=97.6\%$ for both processes, in good agreement with the experimental values.

\begin{figure}[hbt]
    \includegraphics{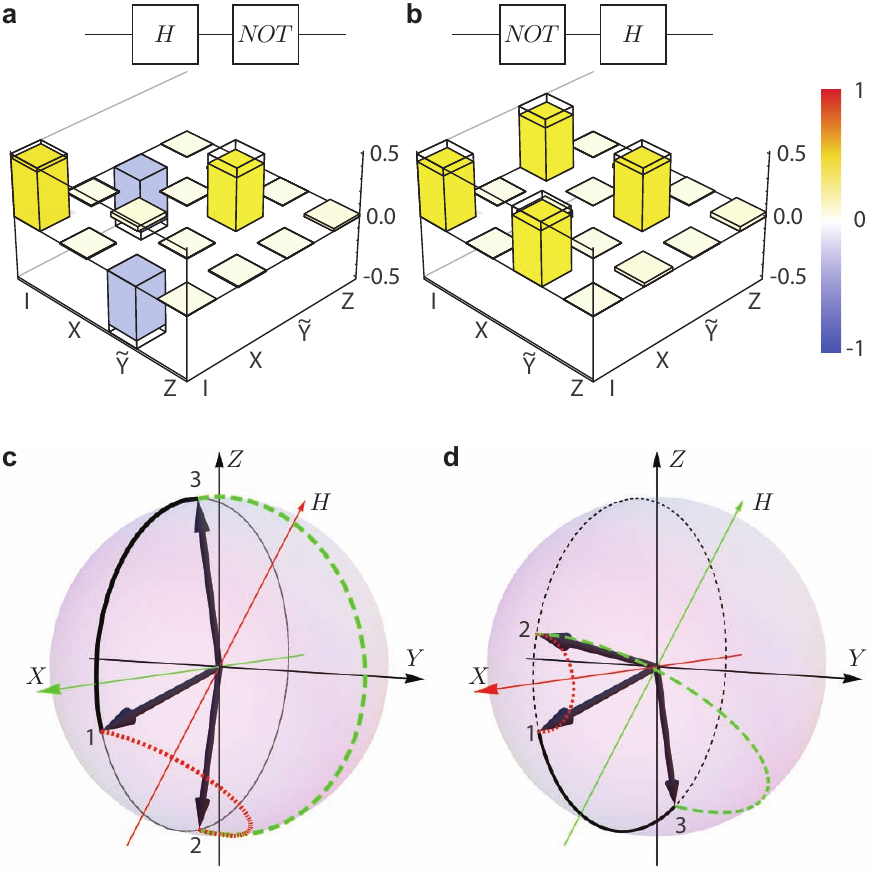}
    \caption{\textbf{Non-commutativity of holonomic gates.} Process matrices for $NOT\cdot H$ (\textbf{a}) and $H\cdot NOT$ (\textbf{b}) gates with fidelities of 95\%. Because of the non-abelian character of the geometric operations the resulting processes are different. This can be visualized on the Bloch sphere by two rotations around $X$ and $H$ axes \textbf{c-d}. The initial state 1 of the system is rotated around the red axis first (red dotted traces) toward the intermediate state 2 and then rotated around the other axes (green axis and dashed lines) to the final state 3. The resulting combined rotations are shown by thick black lines.}\label{fig4}
\end{figure}

To explicitly show that different loops of the state vector in the  Hilbert space result in non-commuting quantum gates, we sequentially apply the geometric Hadamard and the $NOT$ gate in alternating order. The non-abelian character of the operation yields either the operation  $NOT\cdot H = -(\ii\sigma_y+\mathds{1})/\sqrt{2}$ (Fig.\,\ref{fig4}a) or $H\cdot NOT = (\ii\sigma_y-\mathds{1})/\sqrt{2}$ (Fig.\,\ref{fig4}b).
We visualize the operations on the Bloch sphere (Fig.\,\ref{fig4}c-d): $NOT\cdot H$ corresponds to a $\pi$-rotation about the $H$-axis followed by a $\pi$ rotation about the $X$-axis. This is equivalent to a rotation about the $Y$-axis by $\pi/2$.
On the other hand, $H\cdot NOT$ corresponds to a rotation in the opposite direction, which is in clear contrast to the former operation. By concatenation of operations other than the Hadamard and the $NOT$ operations, rotations about arbitrary axes corresponding to a representation of the complete $SU(2)$ group can be realized. 

With the explicit demonstration of two non-commuting purely geometric gates, we have shown the universal character of the realized non-adiabatic non-abelian  transformation acting on a three level quantum system. A similar scheme applied to two coupled three level systems would complete the universal set of holonomic quantum gates and would allow for performing all-geometric quantum algorithms, which are potentially resilient against noise when short pulses are used
\cite{Johansson2012}. Moreover, holonomic gates demonstrated for superconducting quantum devices, could also be applied to other three-level systems with similar energy level structure.

\subsection{Methods: Process tomography}
In order to characterize the gates, we perform full process tomography on the three-level system and reconstruct the process matrix $\chi_{\rm exp}$ using a maximum likelihood procedure \cite{Jezek2003}. Any $SU(3)$ operator acting on a three-level system can be decomposed into nine orthogonal basis operators. As a full set of basis operators we choose $\{I_{01}$, $\sigma_{01}^x$, $-\ii\sigma_{01}^y$, $\sigma_{01}^z$, $\sigma_{0e}^x$, $-\ii\sigma_{0e}^y$, $\sigma_{1e}^x$, $-\ii\sigma_{1e}^y$,$E\}$, where $\sigma_{ij}$ are Pauli operators acting on the levels $i$ and $j$, $I_{01}=\ket{0}\bra{0}+\ket{1}\bra{1}$ and $E=\ket{e}\bra{e}$. The process is fully determined by its action on a complete set of nine input states $\{\ket{0}$, $\ket{e}$, $\ket{1}$, $({\ket{0}+\ket{e}})/{\sqrt{2}}$, $({\ket{0}+\ii\ket{e}})/{\sqrt{2}}$, $({\ket{0}+\ket{1}})/{\sqrt{2}}$, $({\ket{0}+\ii\ket{1}})/{\sqrt{2}}$, $({\ket{e}+\ket{1}})/{\sqrt{2}}$, $({\ket{e}+\ii\ket{1}})/{\sqrt{2}}\}$. These states are prepared by sequentially applying identity, $\pi-$, and $\pi/2-$operations ($I$, $R_x(\pi)$, $R_x(\pi/2)$, $R_y(\pi/2)$) on the $\ket{0}\leftrightarrow\ket{e}$ and $\ket{e}\leftrightarrow\ket{1}$ transitions. After applying the geometric operation to the input states, we perform full state tomography on the respective output states \cite{Bianchetti2010}. The length of a typical sequence is 208\,ns (five 40\,ns pulses with 2\,ns separation). We calibrate the $\pi$-pulses on the $\ket{0}\leftrightarrow\ket{e}$ and the $\ket{e}\leftrightarrow\ket{1}$ transitions by measuring Rabi oscillations between the corresponding states. From the recorded $9^2=81$ measurements, the process matrix $\chi_{\rm exp}$ is reconstructed. The process is compared to the ideal one (\ref{idealproc}) by calculating its fidelity as  $F=\tr\left(\chi_{\rm exp}\chi_{\rm th}\right)$. Since the $\ket{e}$ state serves only as an auxiliary state, we present only the reduced density matrix $\tilde\chi$, which describes the processes involving the $\ket{0}$ and $\ket{1}$ states and omits any operators acting on the $\ket{e}$ state. The set of basis operators is thus given by $\{I,X,\tilde{Y},Z\}=\{I_{01},\sigma_{01}^x,-\ii\sigma_{01}^y,\sigma_{01}^z\}$.



\vspace{0.75cm}
\textbf{Acknowledgement}

\vspace{.2cm}
We are grateful to Erik Sj\"oqvist for fruitful discussions. Supported by the EU project GEOMDISS, the
Austrian Science Foundation (S. F.), and the Swiss National Science Foundation (SNSF).

\vspace{0.5cm}
\textbf{Author Contributions}

\vspace{.2cm}
A.A.A.~\& S.F.~developed the idea for the experiment; A.A.A.~performed the measurements
and analysed the data; J.M.F.~designed and fabricated the sample, J.M.F., K.J., M.P.~\& S.B.~contributed to the
experiment; A.A.A.~\& S.F.~wrote the manuscript; A.W. and S.F. planned the project; all authors commented on the manuscript.

\end{document}